\documentclass[a4paper,12pt]{article}
\pdfoutput=1
\usepackage{amsmath}
\usepackage{amsfonts}
\usepackage{amssymb}
\usepackage{graphicx}
\usepackage{multirow}
\usepackage[small,bf]{caption}
\usepackage{jheppub1}
\usepackage{slashed}
\usepackage{hyperref}                 % For creating hyperlinks in cross
\hypersetup{                            % references
             colorlinks = true,
	     linkcolor = blue,
	     linktocpage = true,
             citecolor = blue,
             urlcolor = blue
           }
\def\hhref#1{\href{http://arxiv.org/abs/#1}{#1}}

\newcommand{\stau}{\tilde{\tau}}

\newcommand{\neut}{\tilde{\chi}^0_1}
\newcommand{\nneut}{\tilde{\chi}^0_2}
\newcommand{\nnneut}{\tilde{\chi}^0_3}
\newcommand{\nnnneut}{\tilde{\chi}^0_4}

\newcommand{\charg}{\tilde{\chi}^+_1}
\newcommand{\ccharg}{\tilde{\chi}^+_2}

\def\be{\begin{equation}}
\def\ee{\end{equation}}

\def\bea{\begin{eqnarray}}
\def\eea{\end{eqnarray}}
\def\bv{\begin{verbatim}}
\def\ev{\end{verbatim}}

\title{Predictions of a Natural SUSY Dark Matter Model for Direct and Indirect
Detection Experiments}
\author[a]{Subhendra Mohanty}
\author[a]{Soumya Rao}
\author[b]{D.P. Roy}
\affiliation[a]{Physical Research Laboratory,\\ Ahmedabad 380009, India}
\affiliation[b]{Homi Bhabha Centre for Science Education,\\Tata Institute of Fundamental
Research,\\Mumbai-400088, India.}
\emailAdd{mohanty@prl.res.in, soumya@prl.res.in, dproy1@gmail.com}

\abstract{
The most natural region of cosmologically compatible dark matter relic density in terms of
low fine-tuning in a minimal supersymmetric standard model with nonuniversal gaugino
masses is the so called bulk annihilation region. We study this region in a simple and
predictive SUSY-GUT model of nonuniversal gaugino masses, where the latter transform as a
combination of singlet plus a nonsinglet representation of the GUT group SU(5). The model
prediction for the direct dark matter detection rates is well below the present CDMS and
XENON100 limits, but within the reach of a future 1Ton XENON experiment. The most
interesting and robust model prediction is an indirect detection signal of hard positron
events, which resembles closely the shape of the observed positron spectrum from the
PAMELA experiment.}

\begin{document}

\maketitle
\flushbottom

\section{Introduction}

The most phenomenologically attractive feature of supersymmetry and in particular the
minimal supersymmetric standard model (MSSM) is that it offers a natural candidate for
dark matter in terms of the lightest superparticle (LSP) \cite{1.}. Astrophysical
constraints on dark matter requires it to be a neutral and colourless particle, while
direct detection experiments strongly disfavour a sneutrino LSP. That makes the lightest
neutralino state ${\tilde{\chi}}^0_1$ (abbreviated as $\chi$) the favoured candidate for
dark matter in the MSSM.  In the constrained version of the model (CMSSM), corresponding
to universal gaugino and scalar masses at the GUT scale, the lightest neutralino state is
dominantly a bino over most of the parameter space. Since the bino carries no gauge
charge, its main annihilation mechanism is via sfermion exchange in the t-channel. This is
usually called the bulk annihilation process; and the region of parameter space giving
cosmologically compatible dark matter relic density via this mechanism is called the bulk
region. It provides the most natural solution to the dark matter problem, in the sense
that the desired dark matter relic density can be obtained in this region with practically
no fine-tuning.  However, LEP sets rather stringent lower limits on  the bino LSP as well
as the sfermion masses in the CMSSM, which rules out the parameter space of the bulk
annihilation region \cite{2.}.

The reason for the large bino and sfermion mass limits mentioned above is that the LEP
lower limit on the neutral Higgs boson mass of the MSSM requires a large radiative
correction from top Yukawa coupling, which in turn requires a large stop mass in order to
suppress the canceling contribution from stop exchange. This in turn requires a large
gluino mass contribution to the RGE of stop mass. Since the GUT scale gluino and bino
masses are equal in the CMSSM, this constraint also implies large bino and sfermion masses
at the weak scale via their RGE. Evidently a simple way to make the bulk annihilation
region of the MSSM dark matter compatible with the Higgs mass limit from LEP is to give up
the universality of gaugino masses at the GUT scale; and in particular to assume that the
GUT  scale bino mass is significantly smaller than that of gluino. Then the latter can
ensure the Higgs mass limit from LEP, while the former ensures relatively small bino and
right-handed slepton masses at the weak scale via their RGE, as required for the bulk
annihilation region. Moreover there are simple and well motivated models for nonuniversal
gaugino masses at the GUT scale, where one assumes that the latter get contributions from
SUSY breaking superfields belonging to the nonsinglet representations of the GUT group
\cite{3.,3.5}. One can combine these two observations to construct simple and predictive
nonuniversal gaugino mass models, which provide a natural solution to the dark matter
relic density while satisfying all the LEP constraints.

The issue of naturalness and fine-tuning involved in achieving the right dark matter relic
density \cite{4.} was investigated in \cite{5a,5b,6.} for a generic MSSM with nonuniversal
gaugino masses. Assuming the usual measure of fine-tuning,

\begin{equation}
	\Delta_a^\Omega=\frac{\partial\ln (\Omega_{CDM}h^2)}{\partial\ln
	(a)}\qquad\&\qquad \Delta^\Omega=max\left(\Delta_a^\Omega\right), \label{relic}
\end{equation}
where $a$ refers to the input parameters of the model \cite{5a}, it was found that
$\Delta^\Omega\sim 1$  over the bulk region. This means there is practically no
fine-tuning involved in achieving the desired dark matter relic density over the bulk
region. In fact over most of this region $\Delta^\Omega$ was found to be slightly less
than 1, for which the authors called the bulk region 'supernatural' for achieving the
desired dark matter relic density. In contrast all the other regions of right dark matter
relic density like the stau-coannihilation, the resonant-annihilation and the focus-point
regions had 1-2 orders of magnitude higher values of this fine-tuning measure.  Of course
one has to pay the usual fine-tuning price for radiative EW symmetry breaking,
$\Delta^{EW}\sim 100$, for the bulk annihilation region like the other DM relic density compatible regions
of the MSSM. However, a quantitative evaluation of this fine-tuning parameter in
\cite{6.} shows, that the bulk annihilation region has one of the lowest $\Delta^{EW}$ amongst all the DM
relic density compatible regions of the MSSM. Thus the low value of $\Delta^\Omega$ is
achieved here without any additional cost to the $\Delta^{EW}$.

Subsequently this issue was investigated in a set of simple and predictive nonuniversal
gaugino mass models, where the GUT scale gaugino masses are assumed to get contributions
from a combination of two SUSY breaking superfields belonging to singlet and nonsinglet
representations of the GUT group SU(5) \cite{7.} - i.e. the combinations 1+24, 1+75 and
1+200.  In each case one could access the bulk region with $\Delta^\Omega \sim 1$,
implying practically no fine-tuning required to achieve the right dark matter relic
density.   In the present work we have investigated the signatures  of this set of natural
SUSY dark matter models for direct and indirect detection experiments. In section 2, we
summarize the essential ingredients of the model. In section 3, we present some
representative SUSY mass spectra of this model and briefly comment on their implications
for the signatures of the model at LHC. Then we present the model predictions for direct
and indirect dark matter detection experiments in sections 4 and 5 respectively. In
particular we shall see in section 5 that the model predicts a hard positron spectrum like
that reported by the PAMELA experiment \cite{8.}, though it cannot account for the
required boost factor in the rate. We conclude with a brief summary of our results in
section 6.

\section{Nonuniversality of Gaugino Masses in SU(5) GUT}

The above set of models is based on the assumption that SUSY is broken by a combination of
two superfields belonging to singlet and a nonsinglet representation of the simplest GUT
group SU(5) \cite{3.}. The gauge kinetic function responsible for the GUT scale gaugino
masses originates from the vacuum expectation value of the F term of a chiral superfield
$\Omega$ responsible for SUSY breaking,

\begin{equation}
	\frac{\langle F_\Omega\rangle_{ij}}{M_{Planck}}\lambda_i\lambda_j,
	\label{fterm}
\end{equation}
where $\lambda_{1,2,3}$ are the U(1), SU(2), SU(3) gaugino fields - bino, wino and gluino.
Since the gauginos belong to the adjoint representation of the GUT group SU(5), $\Omega$
and $F_\Omega$ can belong to any of the irreducible representations appearing in their
symmetric product,

\begin{equation}
	\left( 24\times 24 \right)_{sym}=1+24+75+200.
	\label{sympro}
\end{equation}
Thus, the GUT scale gaugino masses for a given representation of the SUSY breaking
superfield are determined in terms of one mass parameter by

\begin{equation}
	M_{1,2,3}^G=C_{1,2,3}^nm_{1/2}^n
	\label{mgut}
\end{equation}
where

\begin{equation}
	C_{1,2,3}^{1}=(1,1,1),\; C_{1,2,3}^{24}=(-1,-3,2),\; C_{1,2,3}^{75}=(-5,3,1),\;
	C_{1,2,3}^{200}=(10,2,1).
	\label{cn}
\end{equation}

The CMSSM assumes $\Omega$ to be a singlet, leading to universal gaugino masses at the GUT
scale.  On the other hand, any of the nonsinglet representations for $\Omega$ would imply
nonuniversal gaugino masses via eqs (4) and (5). These nonuniversal gaugino mass models
are known to be consistent with the observed universality of gauge couplings at the GUT
scale \cite{3.,9.}, with $\alpha^G\simeq 1/25$. The phenomenology of these models have
been widely studied \cite{10.}. Note that each of these nonuniversal gaugino mass models
is as predictive as the CMSSM. However, none of them can evade the above mentioned LEP
constraint to access the bulk region . This can be achieved by assuming SUSY breaking via
a combination of a singlet and a nonsinglet superfields \cite{7.}, where the GUT scale
gaugino masses are given in terms of two mass parameters,

\begin{equation}
	M_{1,2,3}^G=C_{1,2,3}^1m_{1/2}^1+C_{1,2,3}^lm_{1/2}^l\quad\&\quad l=24,75
	\mbox{ or } 200.
	\label{mgut2}
\end{equation}
Then the weak scale superparticle masses are given in terms of these gaugino masses and
the universal scalar mass parameter $m_0$ via the RGE. In particular the gaugino masses
evolve like the corresponding gauge couplings at the one-loop level of the RGE, i.e.

\begin{align}\notag
	M_1=\left( \frac{\alpha_1}{\alpha_G} \right)M_1^G\simeq \left(\frac{25}{60}\right)
	C_1^nm_{1/2}^n\\\notag
	M_2=\left( \frac{\alpha_2}{\alpha_G} \right)M_2^G\simeq \left(\frac{25}{30}\right)
	C_2^nm_{1/2}^n\\
	M_3=\left( \frac{\alpha_3}{\alpha_G} \right)M_3^G\simeq \left(\frac{25}{9}\right)
	C_3^nm_{1/2}^n
	\label{rge}
\end{align}
The Higgsino mass parameter $\mu$ is obtained from the electroweak symmetry breaking
condition along with the one-loop RGE for the Higgs scalar mass, i.e.

\begin{equation}
	\mu^2+\frac{M_Z^2}{2}\simeq -m_{H_u}^2\simeq -0.1m_0^2+2.1{(M_3^G)}^2-0.22{(M_2^G)}^2
	+0.19M_2^GM_3^G
	\label{murge}
\end{equation}
neglecting the contribution from the GUT scale trilinear coupling term $A_0$ \cite{11.}.
The numerical coefficients on the right correspond to a representative value of
$\tan\beta=10$; but they show only mild variations over the moderate $\tan\beta$ region.
Although we shall be evaluating the weak scale superparticle masses using the two-loop RGE
code {\tt SuSpect}\cite{12.}, the approximate formulae (7) and (8) will be useful in
understanding some essential features of the results.

\section{The SUSY Spectra at the Weak Scale}

It was shown in the Fig 4 of ref \cite{6.} that the bulk region extends over the parameter
range

\begin{equation}
	M_1^G=150-250\mbox{ GeV}\;;\quad m_0=50-80\mbox{ GeV}
	\label{bulk}
\end{equation}
with a mild anti-correlation between the two parameters. This is because the main
annihilation process for the bino LSP pair is via right-handed slepton exchange

\begin{equation}
	\chi\chi\stackrel{\tilde{l}_R}{\longrightarrow}\bar{l}l
	\label{annbino}
\end{equation}
and the bino mass is determined by $M_1^G$ via the RGE (7), while the right-handed slepton
mass is determined via its RGE by $M_1^G$ and $m_0$ with a mild anti-correlation between
the two parameters. Therefore we have chosen to use two set of input parameters

\begin{equation}
	M_1^{G}=200\mbox{ GeV},\;m_0=70\mbox{ GeV}\quad\&\quad M_1^G=250\mbox{
	GeV},\;m_0=67\mbox{ GeV}
	\label{input}
\end{equation}
to represent the centre and the upper edge of the bulk region. The latter set predicts a
relatively hard positron spectrum for the indirect detection signal similar to that of the
PAMELA experiment as we shall see in section 5. For the second gaugino mass parameter we
have chosen to use $M_3^G$ as input, since it makes the dominant contribution to the weak
scale gluino and squark masses as well as the corresponding Higgsino mass of eq (8). The
remaining gaugino mass $M_2^G$ is then determined in terms of  these $M_1^G$ and $M_3^G$
using eqs (4-6). Using these GUT scale gaugino masses along with the scalar mass $m_0$ as
inputs to the RGE code {\tt SuSpect} \cite{12.}, we have evaluated the weak scale SUSY
spectra for a representative value of $\tan\beta=10$, where we have neglected the
contribution from the GUT scale trilinear coupling term $A_0$.

  \begin{table}
    \begin{center}
      $M_1^G=200$ GeV, $M_3^G=800$ GeV, $m_0=70$ GeV\\
      \begin{tabular}{|l|l|l|}
	\hline
	\multirow{2}{*}{Particle} & \multicolumn{2}{|c|}{Mass (GeV)}\\
	\cline{2-3}
	& (1+75) model & (1+200) model\\
	\hline
	$\neut$ (bino)               & 78.4 & 78.0 \\
	$\nneut$ (wino)              & 783 & 582 \\
	$\nnneut$ (higgsino)         & 929 & 970 \\
	$\nnnneut$ (higgsino)        & 954 & 979 \\
	$\charg$ (wino)              & 783 & 582 \\
	$\ccharg$ (higgsino)         & 954 & 979 \\
	\hline
	$M_1$                   & 79.9 & 79.7 \\
	$M_2$                   & 791 & 574 \\
	$M_3$                   & 1718 & 1723 \\
	$\mu$                        & 925 & 965 \\
	\hline
	$\tilde{g}$                  & 1766 & 1766 \\
	$\tilde{\tau}_1$             & 86.3 & 90.8 \\
	$\tilde{\tau}_2$             & 637 & 470 \\
	$\tilde{e}_R,\tilde{\mu}_R$  & 108 & 107 \\
	$\tilde{e}_L,\tilde{\mu}_L$  & 638 & 470 \\
	$\tilde{t}_1$                & 1219 & 1251 \\
	$\tilde{t}_2$                & 1544 & 1506 \\
	$\tilde{b}_1$                & 1513 & 1479 \\
	$\tilde{b}_2$                & 1531 & 1528 \\
	$\tilde{q}_{1,2,R}$          & $\sim 1527$ & $\sim 1533$ \\
	$\tilde{q}_{1,2,L}$          & $\sim 1643$ & $\sim 1592$\\
	\hline
      \end{tabular}
    \end{center}
    \caption{The SuSy mass spectrum for the (1+75) and (1+200) models for a $\sim 80$ GeV LSP.
	    We display the hierarchy and flavour of the neutralino and chargino sectors.
	    We also display the values of the neutralino mass parameters for completeness.
	    For the squarks we take a typical squark mass rather than list the full squark
	    spectrum. The exceptions are the 3rd family squarks that we list separately.
	    Finally, the sneutrinos are degenerate with $\tilde{e},\tilde{\mu}_L$.  The
    lightest higgs mass in this case is 119 GeV for both models.}
	    \label{tab:spec80}
  \end{table}

  \begin{table}
    \begin{center}
      $M_1^G=250$ GeV, $M_3^G=800$ GeV, $m_0=67$ GeV\\
      \begin{tabular}{|l|l|l|}
	\hline
	\multirow{2}{*}{Particle} & \multicolumn{2}{|c|}{Mass (GeV)}\\
	\cline{2-3}
	& (1+75) model & (1+200) model\\
	\hline
	$\neut$ (bino)               & 100 & 99.6 \\
	$\nneut$ (wino)              & 772 & 586 \\
	$\nnneut$ (higgsino)         & 933 & 970 \\
	$\nnnneut$ (higgsino)        & 955 & 979 \\
	$\charg$ (wino)              & 772 & 586 \\
	$\ccharg$ (higgsino)         & 955 & 979 \\
	\hline
	$M_1$                   & 102 & 102  \\
	$M_2$                   & 778 & 579 \\
	$M_3$                   & 1718 & 1723\\
	$\mu$                        & 928 & 965 \\
	\hline
	$\tilde{g}$                  & 1766 & 1766 \\
	$\tilde{\tau}_1$             & 100 & 104 \\
	$\tilde{\tau}_2$             & 627 & 474 \\
	$\tilde{e}_R,\tilde{\mu}_R$  & 119 & 119 \\
	$\tilde{e}_L,\tilde{\mu}_L$  & 628 & 474 \\
	$\tilde{t}_1$                & 1221 & 1251 \\
	$\tilde{t}_2$                & 1541 & 1507 \\
	$\tilde{b}_1$                & 1512 & 1480 \\
	$\tilde{b}_2$                & 1529 & 1528 \\
	$\tilde{q}_{1,2,R}$          & $\sim 1528$ & $\sim 1533$  \\
	$\tilde{q}_{1,2,L}$          & $\sim 1640$ & $\sim 1593$  \\
	\hline
      \end{tabular}
    \end{center}
    \caption{The SUSY mass spectrum for the (1+75) and (1+200) models for a 100 GeV
	    LSP.  Once again the lightest higgs mass is 119 GeV.}\label{tab:spec100}
  \end{table}

  \begin{table}
    \begin{center}
      $M_1^G=250$ GeV, $M_3^G=800$ GeV, $m_0=80$ GeV\\
      \begin{tabular}{|l|l|l|}
	\hline
	\multirow{2}{*}{Particle} & \multicolumn{2}{|c|}{Mass (GeV)}\\
	\cline{2-3}
	& (1+75) model & (1+200) model\\
	\hline
	$\neut$ (bino)               & 101 & 101 \\
	$\nneut$ (wino)              & 789 & 593 \\
	$\nnneut$ (higgsino)         & 1197 & 1218 \\
	$\nnnneut$ (higgsino)        & 1206 & 1223 \\
	$\charg$ (wino)              & 789 & 592 \\
	$\ccharg$ (higgsino)         & 1206 & 1223 \\
	\hline
	$M_1$                        & 103 & 103  \\
	$M_2$                        & 780 & 581 \\
	$M_3$                        & 1728 & 1732 \\
	$\mu$                        & 1197 & 1217 \\
	\hline
	$\tilde{g}$                  & 1766 & 1767 \\
	$\tilde{\tau}_1$             & 109 & 111 \\
	$\tilde{\tau}_2$             & 649 & 478 \\
	$\tilde{e}_R,\tilde{\mu}_R$  & 128 & 128 \\
	$\tilde{e}_L,\tilde{\mu}_L$  & 631 & 477 \\
	$\tilde{t}_1$                & 1056 & 1096 \\
	$\tilde{t}_2$                & 1488 & 1455 \\
	$\tilde{b}_1$                & 1459 & 1421 \\
	$\tilde{b}_2$                & 1519 & 1524 \\
	$\tilde{q}_{1,2,R}$          & $\sim 1531$ & $\sim 1536$  \\
	$\tilde{q}_{1,2,L}$          & $\sim 1643$ & $\sim 1597$  \\
	\hline
      \end{tabular}
    \end{center}
    \caption{The SUSY mass spectrum for the 1+75 and 1+200 models for a LSP
	    mass of 100 GeV obtained with $A_t=A_b=-1.3$ Tev and $A_\tau=0$ TeV.  It
	    predicts a light Higgs mass of 122 GeV, which agrees with the reported value
	    of 125 GeV within the model uncertainty of ~ 3 GeV.}\label{tab:specrdh}
  \end{table}

We shall concentrate on the 1+75 and 1+200 models, for which the dominant contributions to
the gaugino masses satisfying the bulk region come from the singlet superfields \cite{7.}.
Tables 1 and 2 show the weak scale SUSY spectra in the 1+75 and 1+200 models for the two
representative points of  the bulk region (eq.11) and $M_3^G=800$ GeV.  It should be noted
here that the predicted $\stau_1$ mass  of Table 1 is still marginally outside the LEP
disallowed region \cite{2.}.  As expected the squark and gluino masses are primarily
determined by $M_3^G$ irrespective of the choice of the nonsinglet representation or the
values of $M_1^G$ and $m_0$.  The mass range of 1500-1700 GeV for sqaurks and gluinos may
be within the range of the current 7-8 TeV run of LHC \cite{atlas} and it is well within
that of the 14 TeV run.
%The mass range of 1500-1700 GeV for the squark and gluinos lie outside the discovery
%limit of 7-8 TeV LHC but well within that of the 14 TeV run. 
While the masses of the wino and the left-handed sleptons depend on the choices of the
nonsinglet representation and the input mass parameters, the small $m_0$ values ensure
that the latter is always lighter. Thus the SUSY cascade decay at LHC is expected to
proceed via the left-handed selectron/smuon or one of the two stau states, leading to a
distinctive SUSY signal containing a hard $e/\mu$ or $\tau$-jet along with the
missing-$E_T$. However, a quantitative analysis of these LHC signatures is beyond the
scope of the present work.

The predicted value of the light Higgs boson mass for the SUSY spectra of Tables 1 and 2
is 119 GeV. It can be increased by a few GeV via stop mixing by using a moderately large
and negative $A_0$ for the squark sector \cite{12a,12b} to bring it closer to the reported
value of about 125 GeV \cite{12c,12d} at the cost of a larger fine-tuning parameter for
EWSB \cite{12e}.  It may be noted here that there is an uncertainty of $\sim 3$ GeV in the
SUSY model prediction of the light Higgs boson mass arising mainly from the
renormalisation scheme dependence along with the experimental uncertainty in top quark
mass \cite{13a,13b,13c,13d,13e,13f}.  In particular the on-shell renormalistion scheme
prediction is higher  by 2-3 GeV relative to that of the MS-bar scheme used in {\tt
SuSpect}. Therefore we have computed the SUSY spectrum analogous to Table 2, but with
$A_0=-1.3$ TeV for the squark sector, which raises the Higgs mass to the acceptable range
of 122 GeV, as shown in Table 3. We see by comparing this with Table 2 that there is very
little difference between the respective SUSY mass spectra except for a modest increase of
the $\mu$ parameter and the resulting higgsino masses.

\section{Prediction for Direct Dark Matter Detection Experiments}

\begin{table}[b!]
	  \centering
	  \begin{tabular}{|c|c|c|c|}
	    \hline
	    \multirow{2}{*}{$M_1^G$} & \multirow{2}{*}{$M_3^G$} &
	    \multicolumn{2}{|c|}{$m_0$}\\\cline{3-4}
	     & & $1+75$ & $1+200$ \\\hline
	    \multirow{3}{*}{$150$} & 600 & 80 & 80\\
	     & 800 & 80 & 80\\
	     & 1000 & 89 & 80\\\hline
	    \multirow{3}{*}{$200$} & 600 & 70 & 70\\
	     & 800 & 70 & 70\\
	     & 1000 & 77 & 70\\\hline
	    \multirow{3}{*}{$250$} & 600 & 60 & 60\\
	     & 800 & 70 & 66\\
	     & 1000 & 83 & 76\\
	    \hline
	  \end{tabular}
	  \caption{The values of $m_0$, $M_1^G$ and $M_3^G$ used for the benchmark points shown
		  in Fig.~\ref{fig:dd}.}
	  \label{tab:scan}
\end{table}

The direct dark matter detection experiments are mainly based on its elastic scattering on
a heavy nucleus like Germanium or Xenon, which is dominated by the spin-independent
$\chi\, p$ scattering contribution mediated by the Higgs boson exchange. Since the Higgs
coupling to the lightest neutralino $\chi$ is proportional to the product of its gaugino
and higgsino components, the direct detection cross-section is predicted to be small for a
bino dominated $\chi$ state. Fig 1 shows the predicted spin-independent $\chi p$
cross-section for the 1+75 and 1+200 models for 3 representative points in the bulk region
listed in Table~\ref{tab:scan} with $M_3^G=600$, 800 and 1000 GeV. We do not show the
prediction for lower values of this parameter since $M_3^G=500$ GeV corresponds to both
squark and gluino masses in the range of 1000 to 1200 GeV, which may have been already
ruled out by the 7 TeV LHC data \cite{13.}. Note that the size of the higgsino component
of $\chi$ goes down with increasing higgsino mass $\mu$, which is primarily determined by
$M_3^G$ via eq (8). Therefore the direct detection cross-section goes down steadily with
increasing $M_3^G$ with very little dependence on the choice of the nonsinglet
representation or the other input parameters.  We show in this figure the current upper
limits on this cross-section from the CDMS \cite{14.} and XENON100 \cite{15.} experiments
along with the projected limit from a future 1 Ton XENON experiment \cite{16.}. The
predicted rates are seen to be well below the current experimental limits.  However, they
are within the reach of 1 Ton XENON experiment.

\begin{figure}[h!]
	\begin{center}
		\includegraphics[width=0.9\textwidth]{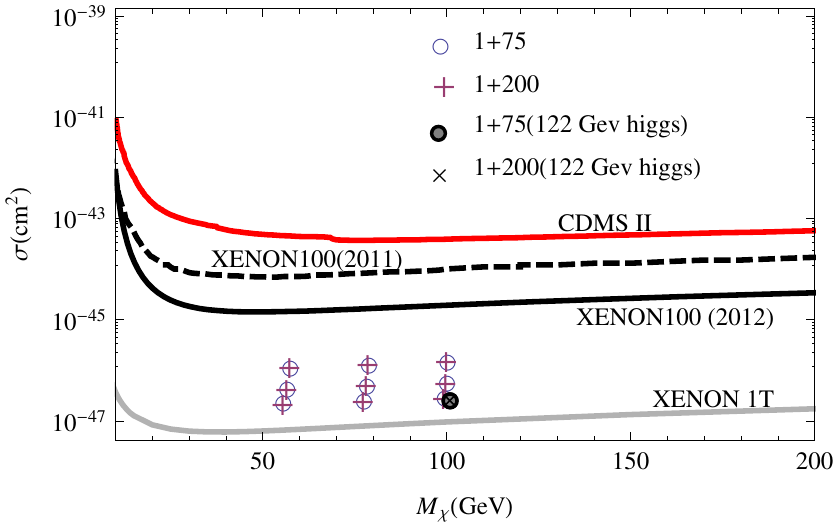}
	\end{center}
	\caption{Plot shows the spin independent (SI) neutralino-proton cross section as a
		function of LSP mass for the 1+75 and 1+200 models.  The value of
		$m_0$, $M_1^G$ and $M_3^G$ used for each point shown in the figure are as
		per Table~\ref{tab:scan}.  With increasing value of $M_3^G$ the cross
		section goes down while $M_1^G$ fixes the LSP mass.  Limits from
		CDMSII\cite{14.} (in red), XENON100\cite{15.} (in black) and the possible
		exclusion limits from the future XENON1T\cite{16.} (in grey) are also
		shown. The present exclusion limit from XENON100 \cite{15.} is down to $2
		\times 10^{-45} cm^2$ ( for DM mass $\sim 55$ GeV) at 90 $\%$ C.L.  We
		also show the direct detection result for the spectrum in
		Table~\ref{tab:specrdh}( filled circle and cross).  This spectrum shows an
		increase in the $\mu$ parameter and higgsino masses compared to the
		spectrum in Table~\ref{tab:spec100}, resulting in a smaller higgsino
		component in the predominantly bino LSP which leads to a slight decrease
		in the direct detection rate.}
	\label{fig:dd}
\end{figure}

\section{Predictions for Indirect Dark Matter Detection Experiments}

The indirect dark matter detection experiments are based on detecting the products of dark
matter pair-annihilation at the present time. Since the dark matter particles are highly
non-relativistic $(v \sim 10^{-3})$, only the s-wave annihilation cross-section is of any
significance at the present time. One can then show from symmetry considerations that for
Majorana particles like the neutralino $\chi$ the cross-section for the annihilation
process (10) is helicity suppressed by a factor of  $(m_l/M_W)^2$. In contrast
cross-section  for the radiative annihilation process

\begin{equation}
	\chi\chi\stackrel{\tilde{l}_R}{\longrightarrow}\bar{l}l\gamma
	\label{radann}
\end{equation}
is only suppressed by a factor of $\alpha$ \cite{17.}. Therefore it provides the dominant
annihilation mechanism at the present time. It can be observed by detecting the  electron/positron,
photon or neutrino (coming from decay of $\mu$ and $\tau$  leptons). A popular indirect
detection experiment is IceCube \cite{18.}, looking for high energy neutrinos coming from
the dark matter pair-annihilation inside the sun. In this case the signal size is
determined by the dark matter capture cross-section by the solar matter, which is mainly
proton; and the main contribution comes from the spin dependent scattering via $Z$ boson
exchange.  Unfortunately, the $Z$ boson coupling to $\chi$ is proportional to the square
of its higgsino component, which is very highly suppressed for the bino dominated dark
matter of our interest. Therefore it offers no viable signal for such experiments.

The most promising signal in this case is provided by the hard positron spectrum coming
from the annihilation process (12). We have evaluated this positron spectrum by computing
the annihilation cross-section for (12) using DarkSUSY \cite{19.}, followed by  the
propagation  of positron using Galprop \cite{20.}.  We have used the isothermal dark
matter density profile\cite{21.} in our computation.  Fig 2 shows the shape of the
predicted positron spectrum relative to electron for a 100 GeV bino dark matter,
corresponding to the SUSY mass spectrum of Table 2. The shape of the observed positron
spectrum from the PAMELA experiment \cite{8.} is also shown for comparison. The shape of
the predicted positron spectrum agrees well with the PAMELA data, with only the prediction
undershooting the last data point by two standard deviations.  One may be tempted to fit
the last data point by increasing the bino dark matter mass to 120-130 GeV.  However, this
will take us into the stau co-annihilation region, which requires significantly higher
fine-tuning than the bulk region. Moreover, the shape of the signal gets flatter with the
increasing DM mass, which increases the overall discrepancy between the predicted spectrum
and the data.  Indeed the positron signal from (12) was already studied in the stau
co-annihilation region in ref \cite{17.}, which compared the PAMELA spectrum with the
model predictions for DM masses of 132 and 233 GeV in its Fig 3. A comparison of that
figure with our present Fig 2  shows an evident deterioration of the overall fit with the
PAMELA spectrum by increasing the dark matter mass from 100 to 132 GeV, which is further
aggravated by increasing the mass further to 233 GeV. The rise of the PAMELA spectrum has
a low threshold of $\sim 20$ GeV, which makes it hard to fit with a dark matter mass
larger than 100 GeV via the annihilation process (12). Therefore an extension of the
positron spectrum beyond 100 GeV from PAMELA or the forthcoming AMS2 \cite{22.} data will
provide a decisive test for this annihilation process.

\begin{figure}[t!]
	\begin{center}
		\includegraphics[width=0.9\textwidth]{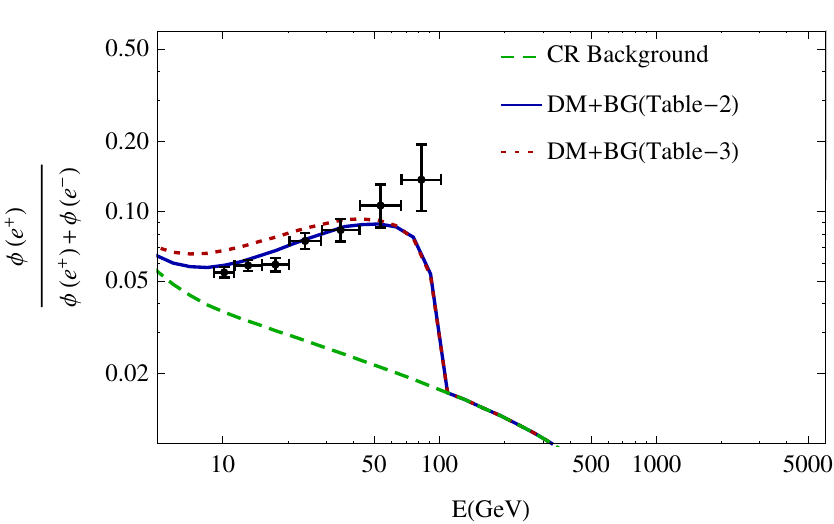}
	\end{center}
	\caption{Ratio of the positron flux to the total ($e^-+e^+$) flux vs energy for a
	100 GeV DM in the 1+75 model , with PAMELA data shown for comparison.  The solid
	line denotes the result for the spectrum in Table \ref{tab:spec100} and the dotted
	line for Table \ref{tab:specrdh}.  The boost in the annihilation cross section is
	taken to be $7000$ and $10000$ respectively.}
	\label{fig:pam}
\end{figure}

Note that the annihilation process (12) does not produce any anti-proton; and hence
predicts no anti-proton excess  over the cosmic ray background in agreement with the
PAMELA data. The main problem in comparing the model prediction with the PAMELA data is
that it requires a large boost factor of  $\sim 7000$ for explaining the size of the
observed positron signal. One can understand this factor as follows. In the most favorable
scenario, where the dark matter is a Dirac particle so that there is no helicity
suppression, the same annihilation process (10) determines the relic density as well as
the size of the PAMELA positron signal \cite{23.}. In this scenario one needs the most
modest boost factor of $\sim 30$. In the present model with Majorana dark matter the
annihilation process (12) responsible for the positron signal is suppressed by a factor of
$\alpha$ relative to process (10) at the freeze-out point, which determines the relic
density.  Therefore for the same relic density the required boost factor for the positron
signal needs to be higher by a factor of $1/\alpha$, which takes it up to $\sim 7000$. It
should be added here that the stau-coannihilation region studied in \cite{17.}, requires
an even larger boost factor of $\sim 30,000$. This is because in that case the pair
annihilation process (10) makes a small contribution relative to stau-coannihilation to
the total annihilation cross-section at freeze-out and the resulting relic density.
Therefore one requires almost an order of magnitude larger boost factor for the
stau-coannihilation region compared to the bulk region. Admittedly, in neither case one
has any explanation for such large boost factors in the SUSY model. Therefore, one needs
to attribute this factor to astrophysical sources like a local population of intermediate
mass black holes leading to spikes in the dark matter density distribution \cite{24.}, or
a nearby dark matter clump \cite{25.}.

\begin{figure}[h!]
	\begin{center}
		\includegraphics[width=0.9\textwidth]{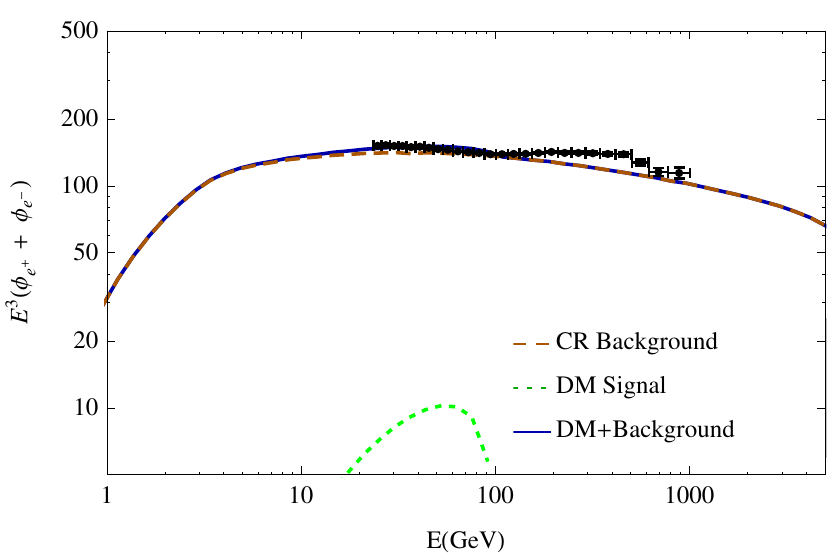}
	\end{center}
	\caption{The total ($e^++e^-$) flux vs energy for a 100 GeV DM with the
	corresponding data from FERMI shown for comparison.  The solid line denotes the
	result for the spectrum in Table \ref{tab:spec100}. The result for Table
	\ref{tab:specrdh} is practically identical to that of Table \ref{tab:spec100}
	shown here.  The boost in the annihilation cross section is taken to be $7000$ for
	Table \ref{tab:spec100} and $10000$ for Table \ref{tab:specrdh}.}
	\label{fig:etot}
\end{figure}

For the reason mentioned earlier, the annihilation process (12) cannot simultaneously
account for the steep rise in the PAMELA positron spectrum as well as the sustained
hardness of the $(e^-+e^+)$ spectrum from the FERMI-LAT data \cite{26.},  spanning over
several hundreds of GeV. Therefore, we assume following \cite{27.}, that the latter can be
accounted for by modifying the cosmic ray propagation parameters within their experimental
uncertainty. Fig 3 shows the predicted $(e^-+e^+)$ spectrum with modified cosmic ray
propagation parameters together with the FERMI-LAT data.  However, we have not tried to
make a detailed fit with the latter by using a larger number of propagation parameters, as
this exercise is not central to the main issue of our paper.  Finally, Fig 4 compares the
predicted $\gamma$ ray spectrum from the annihilation process (12) along with the cosmic
ray background with the FERMI-LAT data \cite{28.}. In this case the signal peak
seems too small to extract from the cosmic ray background contribution to this data.

\begin{figure}[h!]
	\begin{center}
		\includegraphics[width=0.9\textwidth]{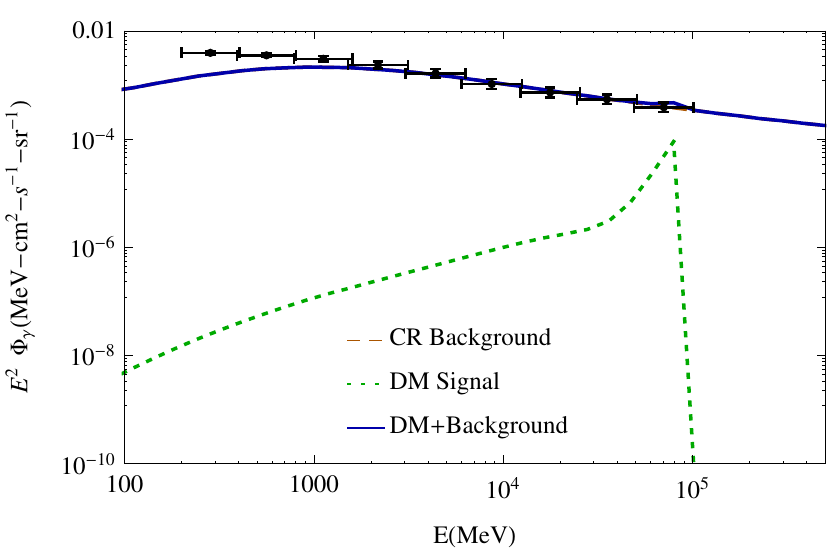}
	\end{center}
	\caption{Diffuse gamma ray flux from 100 GeV DM with data from the FERMI-LAT shown
	for comparison.  The boost in the annihilation cross section is taken to be
	$7000$.}
	\label{fig:ga}
\end{figure}

\section{Conclusions}

Motivated by the observation that the bulk region of dark matter relic density can be
achieved without fine tuning in  models with non-universal gaugino masses at the GUT
scale, specifically those arising from a combination of two SUSY breaking superfields
belonging to singlet and non-singlet representations of SU(5) \cite{7.}, we investigate
the dark matter phenomenology of these models.

We study the signals of the  1+75 and 1+200 models of \cite{7.} in direct and indirect
detection experiments. We scan the parameter space $M_1^G$=150-250 GeV and $m_0$=80-50 GeV
which corresponds to the bulk region and where the bino LSP mass has mass in the range
60-100 GeV. The gluino mass is taken in the range $M_3^G$=600-1000 GeV to evade the bounds
on squark and gluino masses from the 7 TeV LHC data \cite{13.}.

The direct detection cross section for the $\chi\, p$ scattering is small for a primarily
bino LSP because it is mediated by the higgs which couples to the gaugino and Higgsino
components of $\chi$. The recent Xenon100 result with 225 day exposure rules out DM-proton
SI-cross section of up to $2 \times 10^{-45}$cm$^2$ \cite{15.}. The 1+75 and 1+200 models
studied have a lower $\chi\,p$ cross section (Fig 1) and these models are consistent with
direct detection experiments so far. A future Xenon 1T experiment which can probe
$\chi\,p$ cross sections as low as $10^{-47}$ cm$^{2}$ will provide a stringent test of
these models.

The dominant process for indirect detection signal of dark matter is via the s-wave
radiative annihilation, $\chi\chi\stackrel{\tilde{l}_R}{\longrightarrow}\bar{l}l\gamma$
\cite{17.}. This will contribute to the flux of electron/positrons, photons and neutrinos
from $\mu$ and $\tau$ decays. We find that the 100 GeV bino DM can make a significant
contribution to the positron excess observed by PAMELA \cite{8.}. We see from Fig 2 that
the DM annihilation can explain the positron excess (barring the last data point where the
signal is lower than the data within 2-sigma) with a boost factor of $\sim 7000$.  Such a
boost factor may be attributed to astrophysical sources\cite{24.,25.} We do not consider a
higher DM mass as that would require obtaining the required relic density by
stau-coannihilation which would involve a large fine tuning of the parameters at the GUT
scale. Moreover the pair annihilation cross section in the stau-connihilation regime is
smaller so a much larger boost factor $\sim 30,000$ is required \cite{17.} in order to
explain the PAMELA positron signal.  The  measurement of positron flux beyond 100 GeV by
AMS2 \cite{22.} will provide a stringent test of the natural dark matter models \cite{7.}.

\section{Acknowledgement}

This work was initiated during the visit of DPR to the Physical Research Laboratory and
further advanced during the visit of SM to the NIUS (National Initiative in Undergraduate
Science) camp of the Homi Bhabha Centre for Science Education. The work of DPR was partly
supported by the senior scientist fellowship of Indian National Science Academy.

\end{document}